\documentclass[floatfix,twocolumn, 10pt, aps,prd,showpacs,amsmath,amssymb,superscriptaddress]{revtex4-1}
\usepackage{graphicx}
\usepackage{color}
\usepackage{hyperref}
\usepackage[table]{xcolor}
\usepackage{array}
\usepackage{subeqnarray}
\hypersetup{colorlinks=false}
\hypersetup{linktocpage}
\usepackage{soul}
\newcommand{\beq}{\begin{equation}}
\newcommand{\eeq}{\end{equation}}
\newcommand{\beqn}{\begin{eqnarray}}
\newcommand{\eeqn}{\end{eqnarray}}
\newcommand{\nn}{\nonumber}
\definecolor{darkblue}{RGB}{16,78,139}

\usepackage{booktabs}

\begin{document}

\title{Ergoregion instability: the hydrodynamic vortex}

\author{Leandro A. Oliveira}
\email{laoliveira@ufpa.br}
\affiliation{Faculdade de F\'{\i}sica, Universidade  Federal do Par\'a, 66075-110, Bel\'em, Par\'a, Brazil.}
\author{Vitor Cardoso}
\affiliation{Faculdade de F\'{\i}sica, Universidade  Federal do Par\'a, 66075-110, Bel\'em, Par\'a, Brazil.}\affiliation{CENTRA, Departamento de F\'\i sica, Instituto Superior T\'ecnico, Universidade de Lisboa, Av. Rovisco Pais 1, 1049 Lisboa, Portugal.}
\affiliation{Perimeter Institute for Theoretical Physics, Waterloo, Ontario N2L 2Y5, Canada.}

\author{Lu\'is C. B. Crispino}
\email{crispino@ufpa.br}
\affiliation{Faculdade de F\'{\i}sica, Universidade  Federal do Par\'a, 66075-110, Bel\'em, Par\'a, Brazil.}
\date{\today}

\begin{abstract}
Four-dimensional, asymptotically flat spacetimes with an ergoregion but no horizon have been shown to be linearly unstable
against a superradiant-triggered mechanism. This result has wide implications in the search for astrophysically viable
alternatives to black holes, but also in the understanding of black holes and Hawking evaporation. Here we investigate this instability in detail for a particular setup which can be realized in the laboratory: the {\it hydrodynamic vortex},
an effective geometry for sound waves, with ergoregion and without an event horizon.

\end{abstract}
\pacs{
04.70.-s,
04.30.Nk,
43.20.+g,
47.35.Rs 
}
\maketitle

\section{Introduction}

The rotating Kerr family of black holes in general relativity displays a number of 
interesting features. An extreme example is the existence of a static limit, beyond which
every physical observer must co-rotate with the black hole. The static limit defines the outer boundary of the {\it ergoregion},
inside of which observers have access to negative energy trajectories (with respect to static asymptotic observes).
The ergoregion is the chief responsible for a number of interesting effects~\cite{Cardoso:2013krh},
and has a de-stabilizing effect: negative energy states can exit this region with a positive energy, and energy conservation then implies that
left behind are states of even lower negative energy. Fortunately, because the black hole horizon works as a one-way membrane, negative energy states are eventually absorbed and the process is quenched. 

The above reasoning also implies that spacetimes with ergoregions but no horizons are generically unstable, an argument which was made precise by Friedman~\cite{Friedman:1978}. This result has far-reaching consequences, as it renders {\it any} highly rotating, supermassive object unstable,
unless it has a horizon. Thus, ultra-compact alternatives to black holes find here a big obstacle, as the long-term stability
of supermassive objects needs to be explained~\cite{Cardoso:2007az,Chirenti:2008pf}. In gravitation, the ergoregion instability has also been credited with reproducing Hawking radiation in course-grained geometries, as in the fuzzball proposal~\cite{Chowdhury:2007jx,Mathur:2012zp}.

We show here that the ergoregion instability also allows to frame some results about fluid instabilities in the language of curved spacetime. This correspondence arises out of Unruh's seminal work on analogue geometries, showing that sound waves propagating in a background flow are formally equivalent to scalar waves propagating in a curved spacetime with effective metric dictated by the background fluid flow~\cite{Unruh:1980cg,Visser:1997ux,Crispino:proc}. Indeed, the use of analogue spacetimes is also important as a language to understand some of the physics involved in the curved-spacetime setup.

For the first time in the literature, to the best of our knowledge, is presented an investigation of the
time evolutions of the ergoregion instability,
a process which has been fundamental in understanding compact objects, but which
up to now had only been looked at in the frequency domain. We provide a simple, yet
physically realistic acoustic model where such physics can be implemented.
We also note that such configurations had never been previously explored in this context.
Related work includes Ref.~\cite{Slatyer:2005ty} where superradiant amplification factors in a similar geometry are computed, 
with different boundary conditions~\cite{Slatyer_comment}. We do not dwell on amplification factors in this geometry, but rather show that the geometry is unstable.
More specifically, we show that geometries with ergoregion but no horizons are unstable.
Also related to our work, Ref.~\cite{Federici:2005ty}, on the other hand, considers effective geometries with a sonic horizon, and computes amplification factors. 
Our framework has ergoregions but no horizons.

In this paper we provide a direct evidence that the hydrodynamic vortex (an effective spacetime with ergoregion and without an event horizon) 
is unstable under linearized perturbations. We verify explicitly that the appearance of unstable modes in the hydrodynamic vortex
is directly related to the presence of an ergoregion together with the absence of an event horizon.

The remainder of this paper is structured as follows. In Sec. II we describe the spacetime of the hydrodynamic vortex. In Sec. III we study the theory of perturbations of the hydrodynamic vortex using the descriptions in the time and frequency domains. In Sec. IV we obtain the QNM frequencies of the hydrodynamic vortex using three different methods. Specifically, in Subsec. IV.A, using the finite difference (FD) method, we evolve a Gaussian perturbation in the time domain to identify QNM ringing and the profiles of unstable modes. In Subsecs. IV.B and IV.C we use two different frequency-domain methods, the direct integration (DI) and continued fraction (CF) methods, respectively, to obtain the QNM frequencies for fundamental mode $n=0$, as well as higher overtone numbers, with high numerical accuracy. In Sec. V we validate and comment our results comparing the QNM frequencies obtained via FD, DI and CF methods. We conclude with a brief discussion in Sec. VI.

\section{The hydrodynamic vortex}
\label{framework}

We consider an ideal fluid, which is locally irrotational (vorticity free), barotropic, inviscid, and whose background flow velocity $\vec{v}_0$
in spatial $(r,\theta,z)$ coordinates is given by 
\beq
\vec{v}_0= v_r\hat{r} + v_\theta\hat{\theta}+ v_z\hat{z},
\label{velocity}
\eeq
with $v_r = 0$, $v_\theta = v_\theta (r)$ and $v_z = 0$, i.e., 
we assume a two-dimensional purely circulating flow.

Irrotationality implies that the tangential component of the velocity satisfies $v_\theta=C/r$, where $C$ is a constant related to the circulation of the fluid, whereas conservation of angular momentum gives $\rho v_\theta\sim 1/r$. Thus, the background density of the fluid $\rho$ is constant. In turn, this means that the background pressure $P$ and the speed of sound $c$ are constants.

The acoustic spacetime produced by the setup described above is the so-called the 
{\it hydrodynamic vortex}~\cite{drain-footnote}, whose line element is given by \cite{Slatyer:2005ty, Federici:2005ty}
\beq
ds^2=-c^{2}\left(1-\frac{C^2}{c^2r^2} \right) dt^2+dr^2-2Cdtd\theta+r^2d\theta^2+dz^2\,.
\label{vortex}
\eeq
This effective spacetime presents an ergoregion with outer boundary at $r_e = |C|/c$, which coincides with the circle at which the (absolute value of the) background flow velocity equals the speed of sound $c$. Henceforth we set the speed of the sound equal to unity ($c=1$).

\section{Perturbations in a hydrodynamic vortex}
\label{sec-basics}
%
\begin{figure}
   \includegraphics[width=8cm]{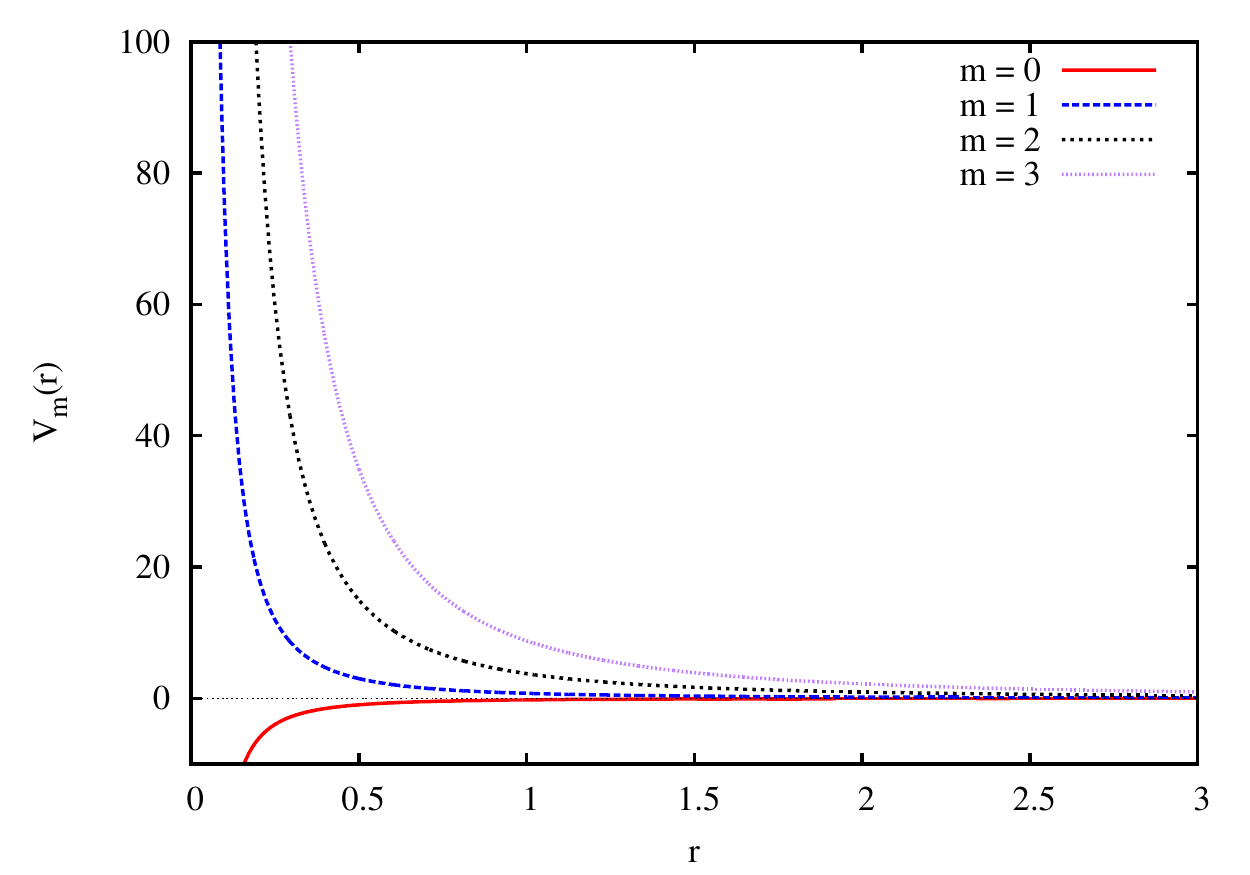}
  \caption{Effective potential $V_m(r)$ for sound wave propagation in the hydrodynamic vortex geometry, as a function of $r$, for azimuthal numbers $m = 0, 1, 2, 3$. }
   \label{pot_Vortex}
  \end{figure}
Consider a background, smooth fluid flow with velocity $\vec{v}_0$, on which small disturbances --sound waves-- propagate, such that
$\vec{v} = \vec{v_0}+\delta \vec{v}$, where $\vec{v}$ is the velocity of the perturbed fluid. Then, linearizing the Navier-Stokes equations around the background flow, it can be shown that small irrotational perturbations $\delta \vec{v} = -\nabla \Phi$ are described by the Klein-Gordon equation \cite{Unruh:1980cg,Visser:1997ux}, namely
\beq
 \Box \Phi = \frac{1}{\sqrt{|g|}} \partial_\mu \left( \sqrt{|g|} g^{\mu \nu} \partial_\nu \Phi \right)=0 , \label{kg}
\eeq
where $g_{\mu \nu}$ is the \emph{effective metric} \eqref{vortex}, with inverse $g^{\mu \nu}$ and determinant $g$. 

Solutions to Eq.~\eqref{kg} are better studied using the cylindrical symmetry of the effective background metric. In particular,
we may decompose the field $\Phi$ in terms of azimuthal modes, namely
\beq
\Phi(t,r,\theta, z) = \frac{1}{\sqrt{r}} \sum_{m=-\infty}^{\infty} \psi_m(t, r) e^{i m \theta}, \label{modesum}
\eeq
where $m$ is an integer to require $\theta-$periodicity.

Inserting Eq.~(\ref{modesum}) into Eq.~(\ref{kg}) leads to the partial differential equation
\beq
\left[-\left(\frac{\partial}{\partial t} + \frac{i C m}{r^2} \right)^2 + \frac{\partial^2}{\partial r^2} - V_m(r) \right] \psi_m(t,r) = 0 ,  \label{waveeq}
\eeq
where the effective potential $V_m(r)$ is
\beq
V_m(r) = \frac{m^2 - 1/4}{r^2}.
\eeq
The effective potential $V_m(r)$ is shown in Fig.~\ref{pot_Vortex} for azimuthal numbers $m=0,1,2,3$. It is negative definite for azimuthal number 
$m=0$, while for azimuthal numbers $m \geq 1$ it is positive definite and decreases monotonically with increasing $r$.

It is also convenient to Fourier transform the function $\psi_m(t,r)$ in Eq.~(\ref{modesum}), and work in the frequency-domain, 
by using the ansatz
\beq
\psi_m(t,r)=u_{\omega m}(r)e^{-i\omega t}\,.
\label{modal}
\eeq
Substituting Eq.~(\ref{modal}) into Eq.~(\ref{waveeq}), we obtain the ordinary differential equation
\begin{eqnarray}
\left[\frac{d^2}{dr^2}+\left( \omega-\frac{Cm}{r^2}\right)^2-V_m(r)\right]u_{\omega m}(r)=0\,,
\label{radial}
\end{eqnarray} 
with irregular singular points at the origin and at infinity. Formally, the Fourier transform requires careful inversion to yield the time-dependent signal $\Phi$~\cite{Leaver:1986gd}. In practice, the response is dominated at intermediate times by the characteristic modes of Eq.~\eqref{radial}, i.e., frequencies $\omega$ for which the solution satisfies the required boundary conditions (see Section~\ref{sec-bc} below). In fact, these modes -- called quasinormal modes (QNMs) -- also allow to understand linear modal stability of the effective geometry~\cite{Berti:2009kk}.

\subsection{Boundary conditions}
\label{sec-bc}
The {\it background} velocity diverges at the origin as $1/r$, signaling a physically singular behavior.
As such, we try to mimic any possible experimental setup by imposing boundary conditions (BCs) at a finite
position $r=r_{\rm min}$. We assume therefore that an infinitely long cylinder of radius $r_{\rm min}$
is placed at the center of our coordinate system. The cylinder is made of a certain material with acoustic impedance $Z$~
\cite{Lax:1948}. 
We consider two different materials corresponding to two limiting values of acoustic impedance 
and therefore to two types of boundary conditions, BC I and BC II, at $r_{\rm min}$. The first one is of Dirichlet type,
\beq
u_{\omega m}(r_{\rm min})=0\,,\quad {\rm BC\,I}\,,\label{BCI}
\eeq
and mimics extremely low-Z materials~\cite{Lax:1948}.

The second boundary condition we consider, BC II, is of Neumann type, and is perhaps easier to mimic experimentally as it 
describes high-Z materials. If we assume a very rigid boundary cylinder, 
of a very high-$Z$ material, the boundary condition amounts to imposing that
$\delta v_r=0$ at $r=r_{\rm min}$, or 
\beq
\frac{d \Phi_{\omega m}}{d r}(r_{\rm min})=0\,,\quad {\rm BC\,II}\,,\label{BCII}
\eeq
which in the frequency domain is equivalent to demanding that $\left(u_{\omega m}/\sqrt{r}\right)'_{r=r_{\rm min}}=0$, 
where primes stand for radial derivatives. 
Both these boundary conditions are realistic and 
whether BC I or BC II holds in practice 
depends on how the entire apparatus is implemented in the laboratory.

At large radial distances we require outgoing Sommerfeld or causal boundary conditions, which in the frequency domain, and given our choice for the Fourier transform, amount to
\beq
u_{\omega m}\left(r\to \infty \right) \sim e^{i\omega r}\,.\label{BC2}
\eeq

Our main task consists therefore in studying solutions of Eqs.~\eqref{waveeq} and \eqref{radial}, subjected to these boundary conditions.
\section{Numerical Methods}
\label{numerical}
We have used the method of lines to integrate Eq.~\eqref{waveeq} in time, but also two different frequency-domain methods to solve Eq.~\eqref{radial}.

\subsection{Finite-difference method (Method of Lines)}
\label{sec-timedomain}
Using the finite-difference (FD) method, we study the evolution of a Gaussian-type disturbance in the time domain~\cite{Rinne:2005df,Witek:2010qc,Witek:2012tr}. Specifically, we used the method of lines, which involves a second-order spatial coordinate discretization and fourth-order Runge-Kutta method to advance in time \cite{Dolan:2011ti,Dolan:2012yt}. We use a Gaussian-type function as initial condition, namely,
\begin{subeqnarray}\label{Gauss}
\slabel{Gauss1} \psi_m(t=0, r) &=& (r-r_{\rm min})^2e^{-\frac{(r -r_0)^2}{2 \sigma^2}},  \\
\slabel{Gauss2} \frac{\partial \psi_m }{\partial t} (t=0, r) &=& 0,
\end{subeqnarray}
where $r_0$ is the position of the center of the peak of the Gaussian function (middle point) and $\sigma$ sets the width of the Gaussian function.

To apply the method of lines in Eq.~(\ref{waveeq}), we discretize the radial coordinate $r \rightarrow r_j = r_{\rm min}+jh$ (for a range $r_{\rm min} \leq r \leq r_{\rm max}$), the wave function $\psi_m(t, r) \rightarrow \psi_j$, and the second-order spatial derivatives
\beqn
\frac{\partial^2 }{\partial r^{2}} \psi_m(t, r) \rightarrow \frac{1}{h^2}\left(\psi_{j+1}-2\psi_{j}+\psi_{j-1}\right) +{\cal{O}}(h^2), \nonumber\\
\eeqn
where $h$ is the resolution of the spatial grid \cite{Rinne:2005df}.
 
With the discretization above, we rewrite Eq.~(\ref{waveeq}) as follows:
\begin{eqnarray}
&&-\frac{d^{2}\psi_{j}}{dt^{2}} - \frac{2i C m}{r_j^2}\frac{d\psi_{j}}{dt} +\left[\frac{C^2 m^2}{r_j^4} - V_m(r_j)\right] \psi_{j}  +\nonumber\\ 
&&\frac{1}{h^2}\left(\psi_{j+1}-2\psi_{j}+\psi_{j-1} \right) = 0. \label{disceq}
\end{eqnarray}
We may transform the second-order differential equation~(\ref{disceq}) in a set of two first-order differential equations using the definition $\zeta_{j}\equiv \frac{d \psi_{j}}{dt}$, namely
\begin{subeqnarray}\label{1ordereqs}
\slabel{1eq}\frac{d \psi_{j}}{dt} &=& \zeta_{j}, \\
\nonumber\\
\slabel{2eq}\frac{d\zeta_{j}}{dt} &=& -\frac{2i C m}{r_j^2}\zeta_{j} + \left[\frac{C^2m^2}{r_j^4} -V_m(r_j)\right] \psi_j+\nonumber\\
 &&\frac{1}{h^2}\left(\psi_{j+1}-2\psi_{j}+\psi_{j-1} \right).
\end{subeqnarray}
We apply the fourth-order Runge-Kutta method in Eqs.~(\ref{1eq}) and~(\ref{2eq}) to evolve the Gaussian-type perturbation (\ref{Gauss}) in the time domain (for a range $0 \leq t \leq t_{\rm max}$) \cite{Dolan:2011ti} to obtain the time profiles for different values of the azimuthal number $m$, circulation $C$ and $r_{\rm min}$. We use the time grid $\tau = h/2$ in our simulations, to ensure the numerical stability of the method of lines and to avoid that instabilities emerge from the method itself.

An exponential growth of the initial values indicates that the solution is unstable. Likewise, stable background geometries give rise to exponentially suppressed sound-wave amplitudes $\psi_m$ and the characteristic QNM frequencies can be estimated by the decay timescale and ringing frequency of the time series.

\subsection{Direct Integration Method}     
\label{sec-direct}
Alternative methods to understand this problem consist in solving directly for the characteristic QNM frequencies of Eq.~(\ref{radial}), using a direct integration (DI) method. 
The DI is specially suited to find unstable modes, since these modes have positive imaginary component and therefore decay exponentially at spatial infinity. The DI is based on the shooting method and numerical root-finding to obtain frequencies in the complex domain \cite{Dolan:2010zza}. We write the outgoing solution at infinity as a generalized power series 
\beq 
u_{\omega m}\left(r\to \infty \right) \sim e^{i\omega r}\sum_{i=0}{\frac{b_i}{ r^i}}\,.
\label{serie2}
\eeq
The series~(\ref{serie2}) and its first-order derivative are then used as boundary conditions to directly integrate Eq.~(\ref{radial}) 
inwards for a range $r_{\rm max} \geq r \geq r_{\rm min}$. 
The QNM frequencies are the roots of this integration procedure that satisfy the appropriate BC I or II, i.e., Eqs.~\eqref{BCI} and \eqref{BCII}, respectively. To find these roots, we use standard root-finding algorithms such as Newton's method.

Alternatively, one can also rewrite Eq.~(\ref{radial}) in terms of $U_{\omega m} \equiv \left(u_{\omega m}/\sqrt{r}\right)'$, as
\beq
\left[ \frac{d^{2}}{dr^{2}}+Q(r)\frac{d}{dr}+R(r)\right]U_{\omega m}=0\,, 
\label{diff_radial}
\eeq 
where
\beqn
&&Q(r) = \frac{5}{r}+\frac{2r(m^2 + 2 C m \omega - 2 \omega^2 r^2)}{C^2 m^2 - m (m + 2 C \omega) r^2 + \omega^2 r^4}, \nn \\ 
&&R(r) = \frac{C^2 m^2}{r^4}+\frac{3-m \left( m + 2\omega C \right) }{r^2} +\omega^2 +\nn \\ 
&& \frac{2(m^2 + 2 C m \omega - 2 \omega^2 r^2)}{C^2 m^2 - m (m + 2 C \omega) r^2 + \omega^2 r^4}\,.\nn
\eeqn
In terms of the function $U_{\omega m}(r)$, BC II corresponds simply to the Dirichlet condition $U_{\omega m}(r_{\rm min})=0$.

\subsection{Continued fraction representation}
\label{sec:ctd-frac}
A powerful alternative to directly integrating Eq.~(\ref{radial}), consists in expressing the problem
as a continued fraction (CF) to find the QNM frequencies \cite{Leaver:1985ax}. We will use a slightly modified
version of the original method (for further details we refer the reader to Appendix B in Ref.~\cite{Pani:2009ss}).

We may define the ``Frobenius'' series in the neighborhood of $r = r_{\rm min}$, 
\begin{equation}
u_{\omega m}(r) = e^{i \omega r} \sum_{n=0} a_n \left( 1 -\frac{ r_{\rm min}}{r} \right)^{n}\,.   \label{cf-ansatz-vt}
\end{equation}
Substituting Eq.~(\ref{cf-ansatz-vt}) into Eq.~(\ref{radial}), we find the following five-term recurrence relation:
\begin{subeqnarray} \label{reco-vt}
&&\alpha_0a_2+\beta_0a_1 +\gamma_0a_0= 0, \\
&&\alpha_1a_3+\beta_1a_2+\gamma_1a_1 +\delta_1a_0 = 0,  \\
&&\alpha_n a_{n+2}+\beta_na_{n+1}+\gamma_n a_{n}+\delta_{n}a_{n-1}+\lambda_{n}a_{n-2} = 0, \nn\\ &&\text{for} \hspace{0.1cm} n \geq 2,
\end{subeqnarray}
where the recurrence coefficients are given by
\begin{subeqnarray}\label{coeff-reco}
\alpha_n &=& 4\left( 1 + n\right)\left(2 + n\right),\\
\beta_n &=& -8 \left( 1 + n\right)  \left( 1 + n -i\omega r_{\rm min}\right),  \\
\gamma_n&=&\left( 1 + 2 n\right) ^2 - 4m^2 \left(1-\frac{C^2}{r_{\rm min}^2}\right) - 8 C m \omega , \\
\delta_{n}&=&  \frac{-8C^2 m^2}{r_{\rm min}^2},\\
\lambda_{n}&=&  \frac{4C^2 m^2}{r_{\rm min}^2}.
\end{subeqnarray}
Using the Gaussian elimination method \cite{Onozawa:1995vu}, we may reduce the 
five-term recurrence relation (\ref{reco-vt}) to a three-term recurrence relation. 
First we reduce Eqs.~\eqref{reco-vt} to a four-term relation by defining,
\begin{subeqnarray}
\alpha'_n &=& \alpha_{n}  ,\\
\beta'_n &=& \beta_{n}, \\
\gamma'_n &=&\gamma_{n}, \\
\delta'_n &=&\delta_{n},  \hspace{0.5cm} \text{for} \hspace{0.1cm} n=0, 1,
\end{subeqnarray}
and
\begin{subeqnarray}
\alpha'_n &=& \alpha_{n}  ,\\
\beta'_n &=& \beta_{n}-\frac{\alpha'_{n-1}\lambda_{n}}{\delta'_{n-1}}, \\
\gamma'_n &=&\gamma_{n}-\frac{\beta'_{n-1}\lambda_{n}}{\delta'_{n-1}}, \\
\delta'_n &=&\delta_{n}-\frac{\gamma'_{n-1}\lambda_{n}}{\delta'_{n-1}}, \hspace{0.5cm} \text{for} \hspace{0.1cm} n \geq 2. 
\end{subeqnarray}

Then, we may use again the Gaussian elimination to reduce further the four-term recurrence relation. We define
\begin{subeqnarray}
\alpha''_n &=& \alpha'_{n},\\
\beta''_n &=& \beta'_{n}, \\
\gamma''_n &=&\gamma'_{n}, \hspace{0.5cm} \text{for} \hspace{0.1cm} n=0,
\end{subeqnarray}
and
\begin{subeqnarray}
\alpha''_n &=& \alpha'_{n}  ,\\
\beta''_n &=& \beta'_{n}-\frac{\alpha''_{n-1}\delta'_{n}}{\gamma''_{n-1}}, \\
\gamma''_n &=&\gamma'_{n}-\frac{\beta''_{n-1}\delta'_{n}}{\gamma''_{n-1}}, \hspace{0.5cm} \text{for} \hspace{0.1cm} n \geq 1,
\end{subeqnarray}
which leads us to the following three-term recurrence relation:
%
\begin{eqnarray}
\alpha''_n a_{n+2}+\beta''_na_{n+1}+\gamma''_n a_{n} = 0, \hspace{0.5cm} \text{for} \hspace{0.1cm} n \geq 0.
\label{reco-vt-3}
\end{eqnarray}

The recurrence coefficients $ \alpha''_n, \beta''_n $ and $ \gamma''_n $ are complex functions that depend on the frequency $ \omega $, the azimuthal number $ m $, the circulation $ C $ and $ r_{\rm min} $. Henceforth, to simplify the notation we omit the double-primes from the recurrence coefficients. It can then be shown from Eq.~(\ref{reco-vt-3}) that the solution one is interested in, called the minimal solution, satisfies the relation
\beq
\frac{a_2}{a_1}= F \,,\quad \text{for} \quad {\rm BC\,I}\,,
\label{fff}
\eeq
and considering $n \rightarrow n-1$ in Eq.~(\ref{reco-vt-3}) we find the following relation
\beq
\frac{a_1}{a_0}= F \,,\quad \text{for} \quad {\rm BC\,II}\,,
\label{fff_n}
\eeq
where $F$ is given by the following continued fraction
\beq
F = -\frac{\gamma_1}{\beta_1-}\frac{\alpha_1\gamma_2}{\beta_2-}\frac{\alpha_2\gamma_3}{\beta_3-}....
\eeq
Thus, both BC I and BC II can be handled with a CF representation by appropriately choosing $a_2/a_1$ and $a_1/a_0$, respectively, in Eqs.~(\ref{fff}) and~(\ref{fff_n}).
It is easy to show that for Dirichlet boundary conditions, BC I [considering $a_0=0$ in Eq.~(\ref{reco-vt-3})], one gets the following equation
\begin{eqnarray}
&& \beta_0-\frac{\alpha_0\gamma_{1}}{\beta_{1}-}\frac{\alpha_{1}\gamma_{2}}{\beta_{2}-}\frac{\alpha_{2}\gamma_{3}}{\beta_{3}-}... = 0. 
\label{highordern-vt}
\end{eqnarray}

For Neumann boundary conditions, BC II, it follows from Eq.~\eqref{cf-ansatz-vt} that we must choose
\begin{equation}
\frac{a_1}{a_0}=\frac{1 - 2i\omega r_{\rm min}}{2}\,,
\label{frac_bcii}
\end{equation}
and solve Eq.~\eqref{fff_n}. Eqs.~(\ref{highordern-vt}) and~(\ref{frac_bcii}) can be solved with standard root-finding algorithms such as Newton's method.

\begin{figure*}
\includegraphics[width=16cm]{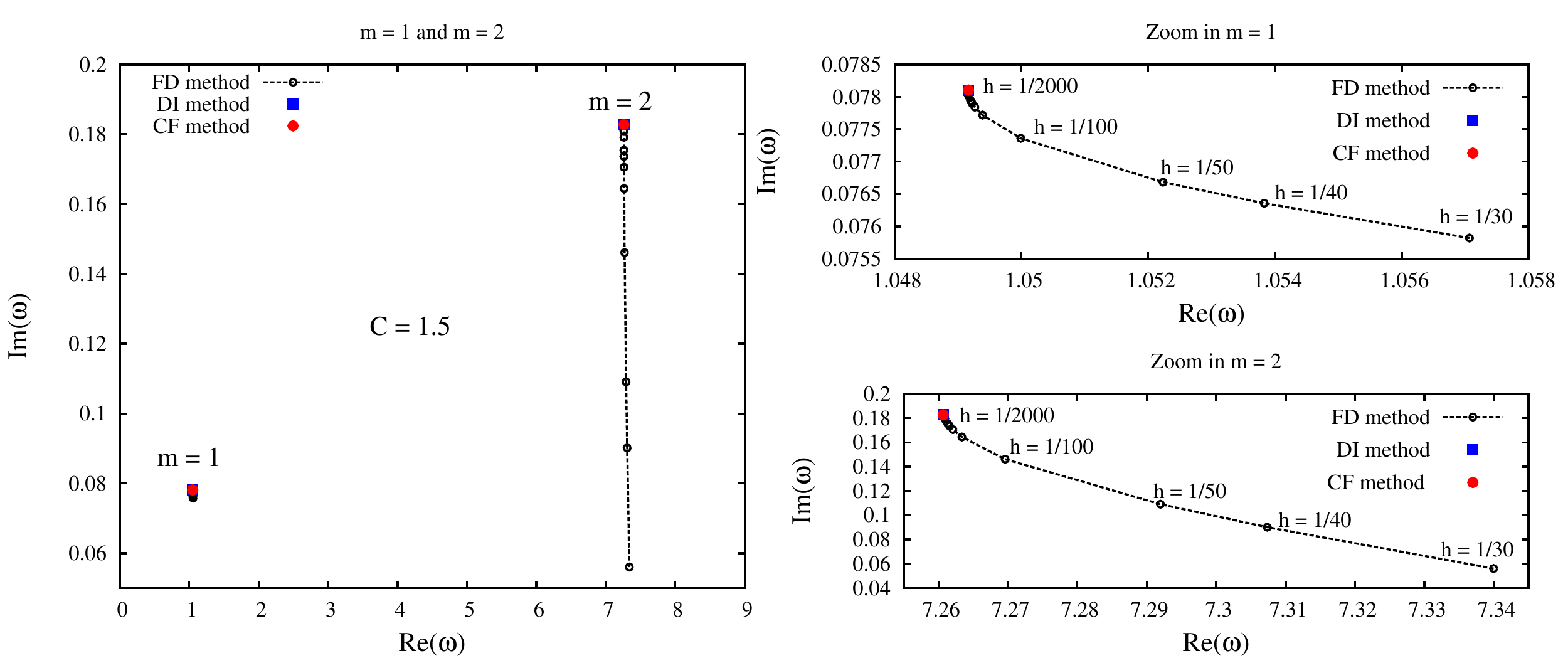}
\caption{Comparison of the characteristic complex QNM frequencies obtained through different methods. We focus on azimuthal numbers $m=1,2$ and circulation $C=1.5$. The finite difference (FD) result is studied as a function of resolution $h=1/30..1/2000$, using a Gaussian-type wavepacket with $r_0 = 2.0$, and $\sigma=0.25$. We impose BC I, given by Eq.~(\ref{BCI}), at $r _{\rm min}=0.3$. At high enough resolution the FD estimates converge towards frequency-domain estimates.}
\label{ring6}
\end{figure*}
%
\section{Results}
\label{results}
%
\begin{table}
\centering \caption{QNM frequencies of the fundamental mode ($n=0$) for azimuthal numbers $m =1, 2$ and circulation $C = 0.5$. We impose BC I for different values of $r_{\rm min}$, estimated via finite difference (FD), direct integration (DI) and continued fraction (CF) methods. The different methods show remarkable agreement. } \vskip 12pt
\begin{tabular}{@{}cc|cccc@{}}
\hline \hline
&&\multicolumn{2}{c}{$m=1$}&\multicolumn{2}{c}{$m=2$}\\ \hline
$r_{\rm min}$    & Method   &$\text{Re}(\omega)$     &$\text{Im}(\omega)$ & $\text{Re}(\omega)$ &$\text{Im}(\omega)$  \\
                 &FD        &$49.758385$            &$2.613961$         &$164.927794$        &$3.380723$ \\
$0.05$           &DI        &$49.745099$            &$2.699752$         &$164.838202$        &$4.323223$  \\
                 &CF        &$49.745100$            &$2.699752$         &$164.838252$        &$4.323224$ \\
                 &FD        &$3.147594$             &$0.233977$         &$21.783365$         &$0.531911$ \\
$0.10$           &DI        &$3.147494$             &$0.234318$         &$21.782052$         &$0.548381$  \\
                 &CF        &$3.147493$             &$0.234318$         &$21.782051$         &$0.548381$ \\
                 &FD        &$-0.486747$            &$-0.117741$         &$3.800125$          &$0.037680$ \\
$0.15$           &DI        &$-0.478036$            &$-0.121902$         &$3.800082$          &$0.038183$ \\
                 &CF        &$-0.478036$            &$-0.121902$         &$3.800081$          &$0.038183$ \\
\hline \hline
\end{tabular}
\label{ringtable1}
\end{table}  

We follow standard conventions of ordering the QNM frequencies by their imaginary part~\cite{Berti:2009kk}.
The fundamental mode is the one with largest imaginary component ${\rm Im}[\omega]$.
Thus, if the mode is unstable (${\rm Im}[\omega]>0$) the fundamental mode corresponds to the smallest instability
timescale, and for stable (${\rm Im}[\omega]<0$) modes it corresponds to the longest-lived mode.

We note that the hydrodynamic vortex presents the following symmetries (these can be obtained from Eq.~(\ref{radial})):
\beq
\omega (m,C) = -\omega^\ast (-m,C) = -\omega^\ast (m,-C) = \omega (-m,-C), \nn
\eeq
where ``${}^\ast$'' denotes complex conjugation.
\subsection{Convergence properties and consistency checks}
%
\begin{figure*}
\includegraphics[width=16cm]{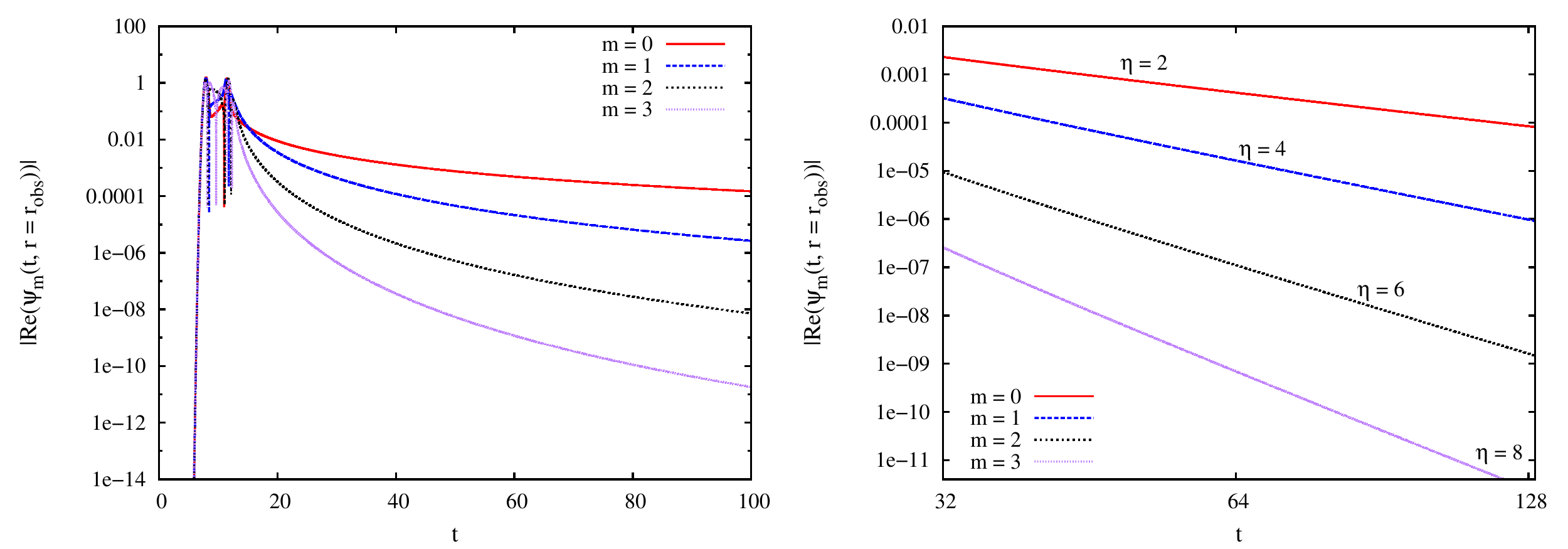}
\caption{Left: Time-domain profiles of $|\text{Re}(\psi_m(t, r))|$ for azimuthal numbers $m = 0, 1, 2, 3$ and circulation $C = 0.0$. Right: Log-log plot illustrating the power-law decay of $|\text{Re}(\psi_m(t, r))|$, as a function of time, for azimuthal numbers $m = 0, 1, 2, 3$,  $r_{\rm min}=0.3$ and circulation $C=0.0$. At late times, the perturbation decays according to an inverse power-law $\psi_m \propto t^{-\eta}$, where $\eta = 2|m|+2$ for an initial condition given by Eq.~(\ref{Gauss}). The late-times tails depend on the leading term of the effective potential at large distances, therefore the decay rate is independent of the circulation $C$ \cite{Berti:2004ju}. Here we extract the signal at $r_{\rm obs}=10$.}
\label{ring1}
\end{figure*}
Before proceeding to look at the overall results, we have tested the convergence of the various methods and their mutual agreement.
Examples are shown in Table~\ref{ringtable1} and Figure~\ref{ring6}.

Figure~\ref{ring6} shows how the estimated QNM frequency from time-domain data changes with resolution $h$ of the spatial grid, and how it approaches the frequency-domain DI and CF methods. For this we evolved in time the Gaussian-type wavepacket~(\ref{Gauss}) with $r_0 = 2.0,\,\sigma = 0.25,\,r _{\rm min}=0.3$ and we extracted the $m=1, 2$ poles of the field $\psi_m$ at $r=10.0$. The grid uses a range $0 \leq t \leq 100$ and $r_{\rm min} \leq r \leq 200$, spatially large enough that the boundary is causally disconnected from the region where the modes are extracted. We then estimated numerically the QNM frequencies from the time-domain profiles of $\psi_m(t, r)$ for different values of resolution $h$. As can be seen from Fig.~~\ref{ring6},
at large enough number of grid points, the results converge and they converge towards values which are in complete agreement with frequency domain methods. Numerical examples of the excellent agreement between the three different methods, imposing BC I, are shown in Table~\ref{ringtable1}.

A further check on our numerical procedure is provided by time evolutions of Gaussian-type wavepackets in ``flat space'' with $C=0$. Backgrounds with zero circulation correspond to a background fluid at rest, such that we are effectively studying sound wave propagation in a trivial background.
The results for such time evolutions are shown in Fig.~\ref{ring1} for different values of the azimuthal number $m$. As expected for Minkowski backgrounds, there are no QNMs, but because wave propagation in flat, odd-dimensional spacetimes~\cite{huygens_comment} develops tails due to the failure of Huygens' principle~\cite{Cardoso:2003jf} the profile is non-trivial. In agreement with a Green's function analysis, the late-times tails decay according to an inverse power-law $\psi_m \propto t^{-\eta}$, where $\eta = 2|m| + 2$ for an initial condition given by Eq.~(\ref{Gauss}) \cite{Dolan:2011ti}.
\subsection{The ergoregion instability}
%
\begin{figure*}
\includegraphics[width=16cm]{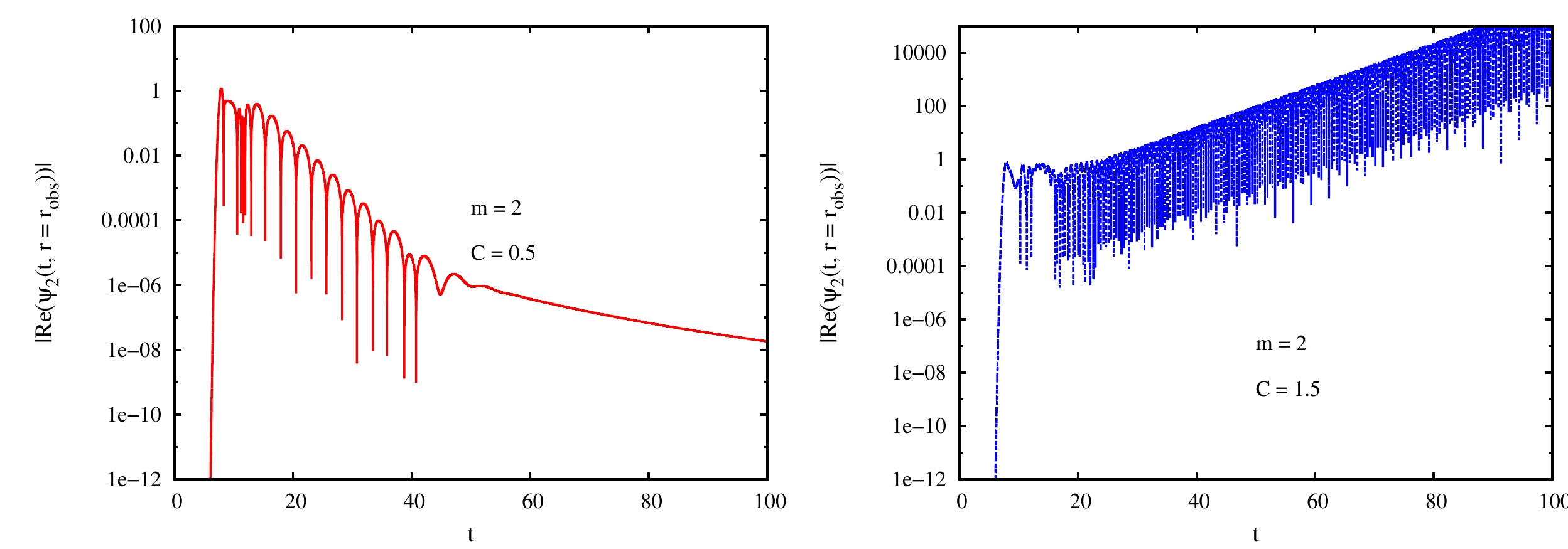}
\caption{Time-domain profiles of $|\text{Re}(\psi_m(t, r))|$ for azimuthal number $m=2$, circulations $C=0.5$ (left)$,\,1.5$ (right), with BC I [cf. Eqs.~\eqref{BCI}] and $r_{\rm min}=0.3$. The wave is extracted at $r_{\rm obs}=10$.}
\label{ring9}
\end{figure*} 
\begin{table}
\centering \caption{QNM frequencies of the fundamental mode ($n=0$) for azimuthal number $m=1$ and circulation $C=0.5$, for 
different values of $r_{\rm min}$.}
\vskip 12pt
\begin{tabular}{@{}ccccccc@{}}
\hline \hline
\multicolumn{1}{c}{} & \multicolumn{2}{c}{BC I}& \multicolumn{2}{c}{BC II}\\ \hline
$r_{\rm min}$        &$\text{Re}(\omega)$ &$\text{Im}(\omega)$ & $\text{Re}(\omega)$ & $\text{Im}(\omega)$\\
0.10                 &$3.147493$              &$0.234318$      &$18.729978$         &$2.351723$  \\
0.15                 &$-0.478036$             &$-0.121902$     &$5.457651$          &$0.739910$ \\
0.30                 &$-0.431205$             &$-0.833382$     &$-0.213441$         &$-0.023529$  \\
\hline \hline
\end{tabular}
\label{ringtable}
\end{table}  
%
As we explained in the Introduction, the physical mechanism triggering the ergoregion instability
is associated with negative-energy states (with respect to asymptotic static observers) which become positive-energy states when they leave the ergoregion. Energy conservation then requires that left behind are modes of energy even more negative (see also Fig.~\ref{fig_snapshots} below).
The only way to quench or kill the instability is to absorb these negative-energy states, as for instance by horizons (and this explains why the Kerr black hole is stable against massless perturbations). This reasoning also anticipates that the choice of different kinds of boundary conditions at $r=r_{\rm min}$ is not fundamental for the existence of the instability. What is fundamental is the location of the surface $r_{\rm min}$. If it lies inside the ergoregion, an instability is expected to be triggered since otherwise we would be cutting the ergoregion away from the physical region under study.

\begin{figure*}
\includegraphics[width=16cm]{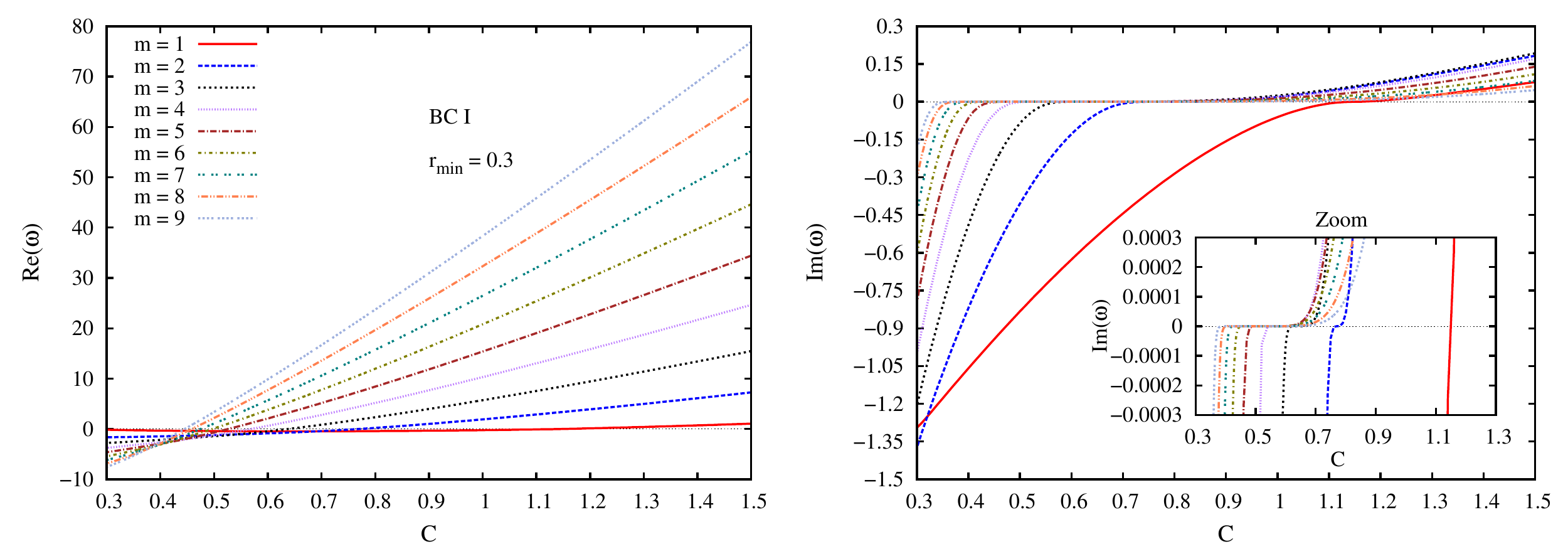}\\
\includegraphics[width=16cm]{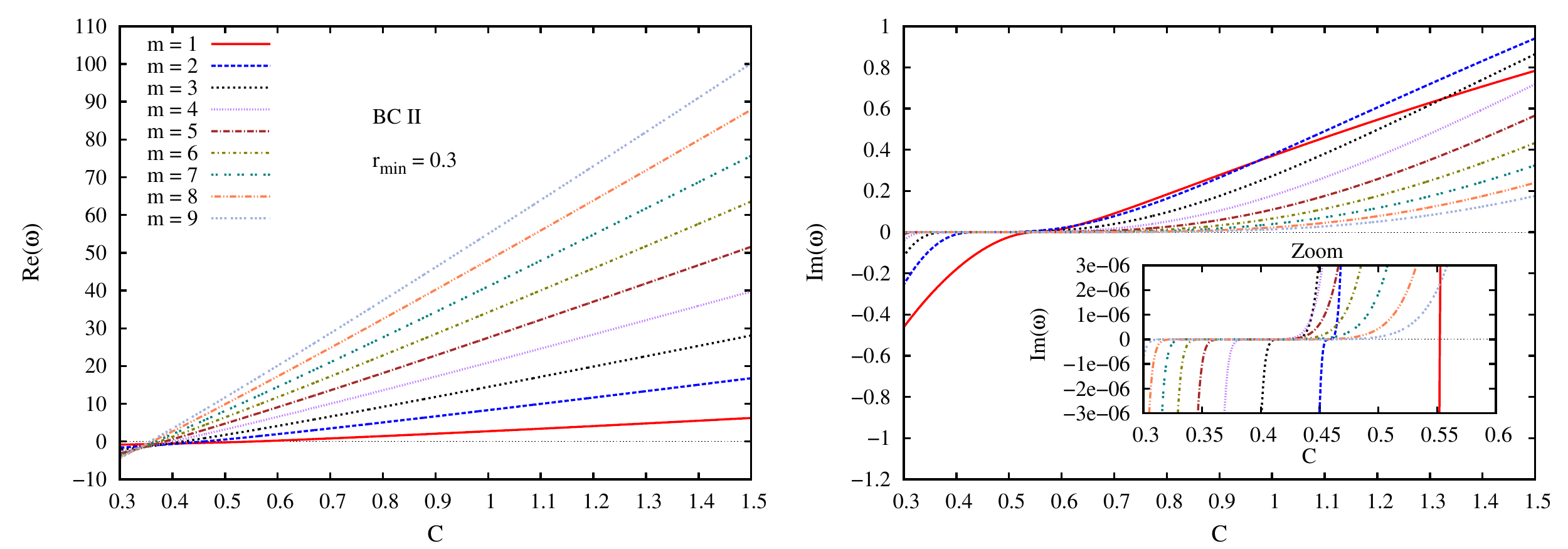}
\caption{Real (left) and imaginary (right) components of the fundamental QNM frequencies, plotted as a function of $C$, for $r_{\rm min}=0.3$ and different values of  $m$.
The top plots correspond to BC I, whereas the bottom plots correspond to BC II. These numbers were extracted with CF methods.
}
\label{fig-stable1}
\end{figure*}

\begin{figure*}
\includegraphics[width=16cm]{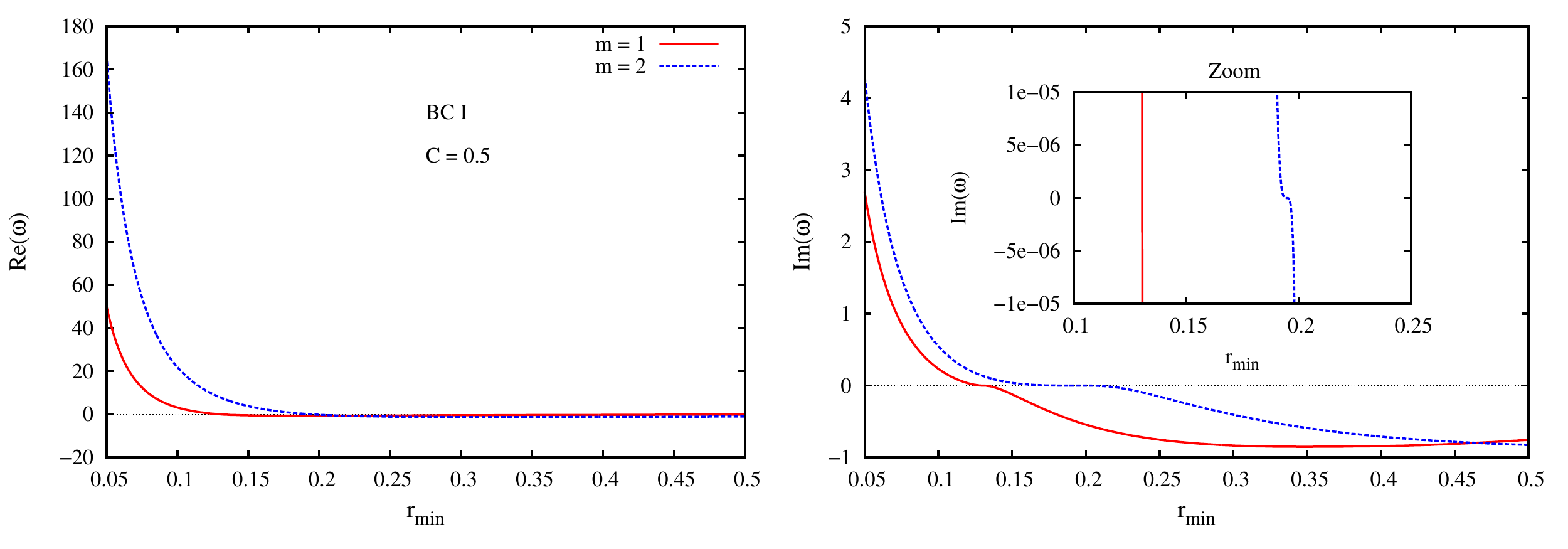}\\
\includegraphics[width=16cm]{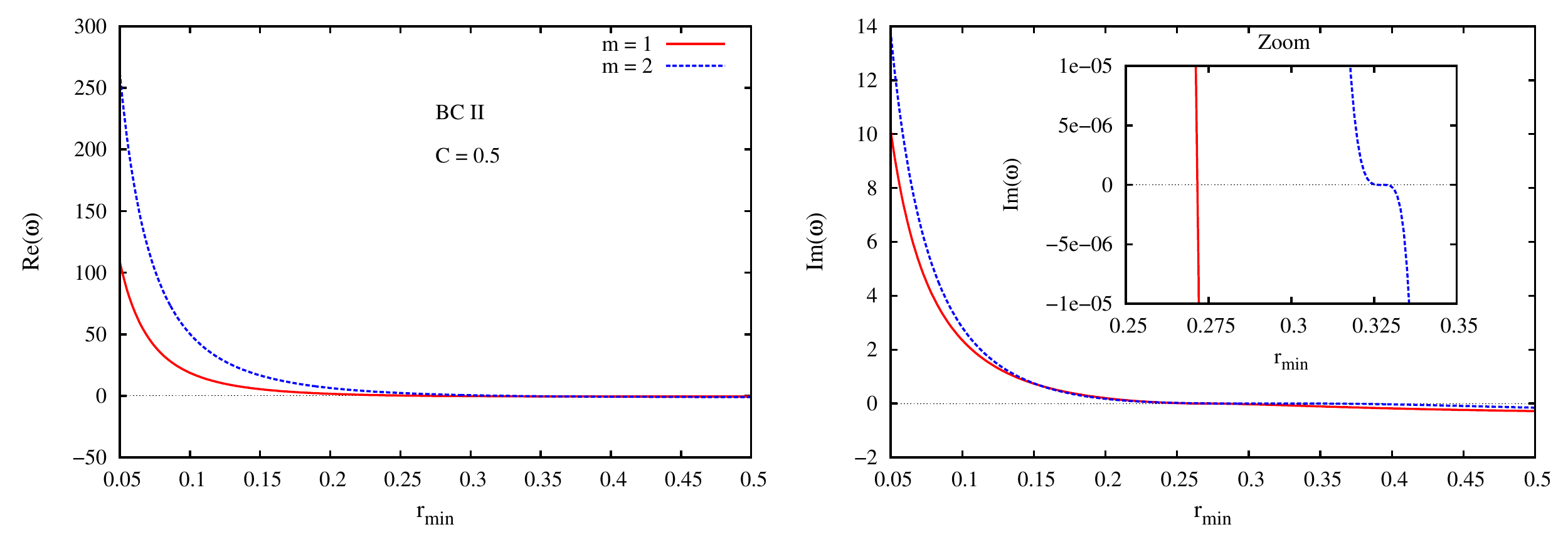}\\
\caption{Real (left) and imaginary (right) components of the fundamental QNM frequencies, plotted as a function of  $r_{\rm min}$, for $C=0.5$ and $m=1,2$.
The top plots correspond to BC I, whereas the bottom plots correspond to BC II. These numbers were extracted with CF methods.
}
\label{fig-stable2}
\end{figure*}

To test these ideas, we considered both BCs I and II, given by Eqs.~\eqref{BCI} and~\eqref{BCII}, respectively,
and $r_{\rm min}=0.3$. For time evolutions, we started with a Gaussian-type wavepacket given by Eq.~\eqref{Gauss}, with 
$r_0=2.0,\,\sigma=0.25$, and a grid $0 \leq t \leq 100$ and $r_{\rm min} \leq r \leq 200$. Our results are summarized in 
Table~\ref{ringtable} and Figs.~\ref{ring9}-\ref{fig-stable2}.
We note that the imaginary part of the QNM frequencies for BC II is bigger than the corresponding one for BC I,
implying that if a QNM is unstable for BC I it will also be unstable for BCII.

Figure~\ref{ring9} shows a typical time-evolution of a Gaussian wavepacket centered at a fixed location, in this case at $r=10$, for different circulations
and $r_{\rm min}=0.3$. The initial perturbation is damped for small enough circulation $C$ and grows exponentially (signaling an instability of the system) for large enough $C$. One seemingly puzzling feature of Fig.~\ref{ring9} is the fact that $m=2$ perturbations for circulation $C=0.5$ are {\it stable}, even though the reflecting BCs are imposed inside the ergoregion. This seems to contradict the claim that the ergoregion instability is a generic feature of horizonless geometries. The key to understand this behavior lies in the fine structure of the ergoregion instability,
which is intrinsically a large-$m$ instability. To dissecate this behavior, we show the details of the instability for higher $m-$modes in 
Figs.~\ref{fig-stable1}. 

In fact, for $r_{\rm min}=0.3$, the large-$m$ threshold of the instability asymptotes to $C=0.3$, as can be seen from Fig.~\ref{fig-stable1}, 
and as anticipated from our discussion. 

Figure~\ref{fig-stable2} shows that the $m=2$ mode is unstable only for values $r_{\rm min}\lesssim 0.19$ for BCI and $r_{\rm min}\lesssim 0.33$ for BCII, but that the instability threshold increases as $m$ increases. Our results also indicate (cf. Figure~\ref{fig-stable1}) that all modes $m>5$ are unstable for $r_{\rm min}=0.3$ and circulation $C=0.5$. Moreover, at fixed inner boundary location $r_{\rm min}$ and fixed $m$ the instability gets stronger for larger $C$, as might also be anticipated. All our numerical results fully support the statement that the presence of an ergoregion without event horizon gives rise to instabilities. 
\begin{figure*}
\includegraphics[width=16cm]{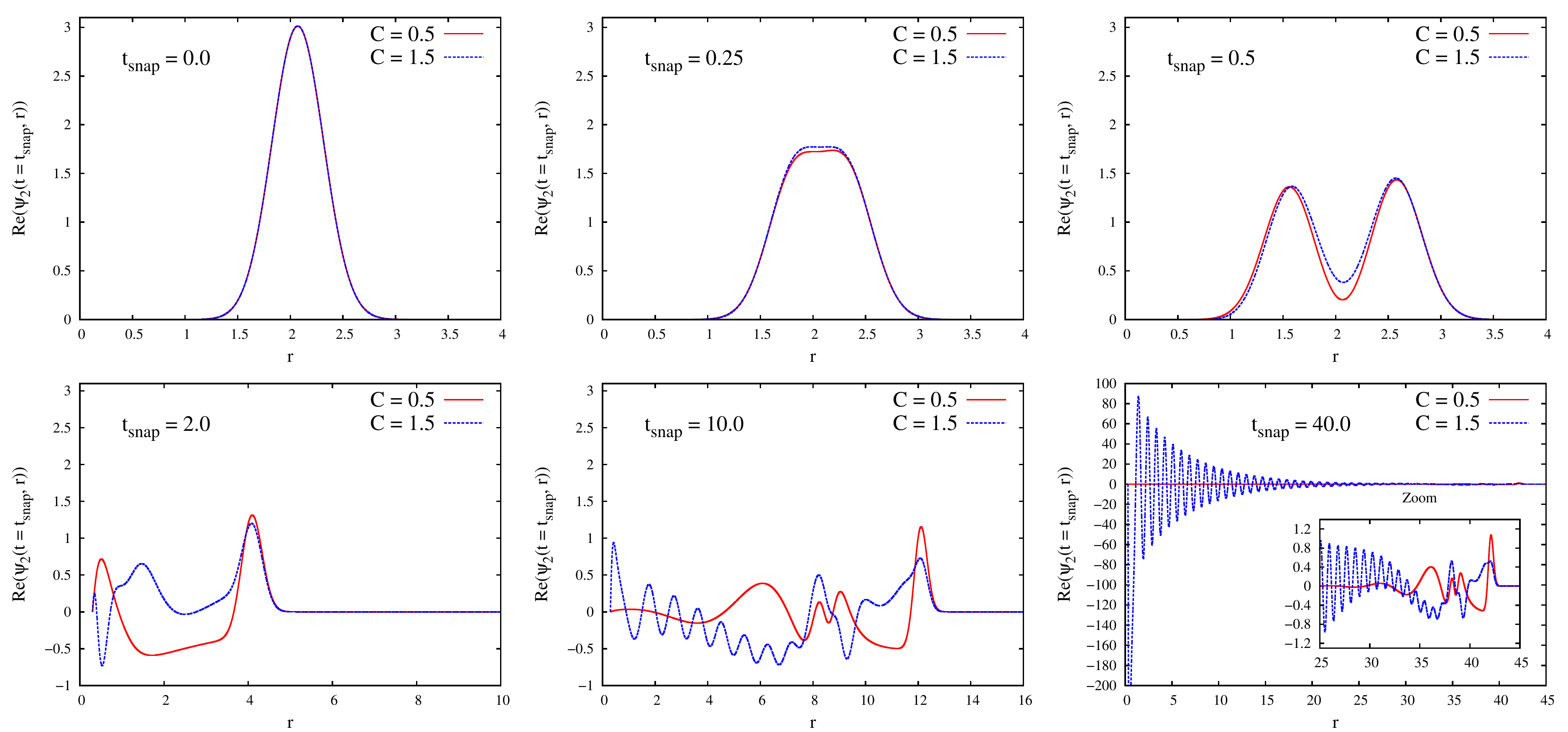}
\caption{Snapshots of the radial profiles of $\text{Re}(\psi_{m}(t, r))$ for azimuthal number $m = 2$, circulations $C=0.5$ (stable case) and $C=1.5$ (unstable case). Boundary conditions BC I are imposed at $r_{\rm min}=0.3$.}
\label{fig_snapshots}
\end{figure*} 

A complementary facet of the instability is shown in snapshots of the evolution, as those depicted in Fig.~\ref{fig_snapshots}. These snapshots compare the evolution of a stable ($C=0.5$) and unstable ($C=1.5$) configuration, both for $m=2$, and show clearly how the instability develops inside the ergoregion and close to the inner boundary at $r_{\rm min}=0.3$. Notice the scale in the last snapshot, and how the field decays in space but grows in time.

\begin{figure*}
\includegraphics[width=16cm]{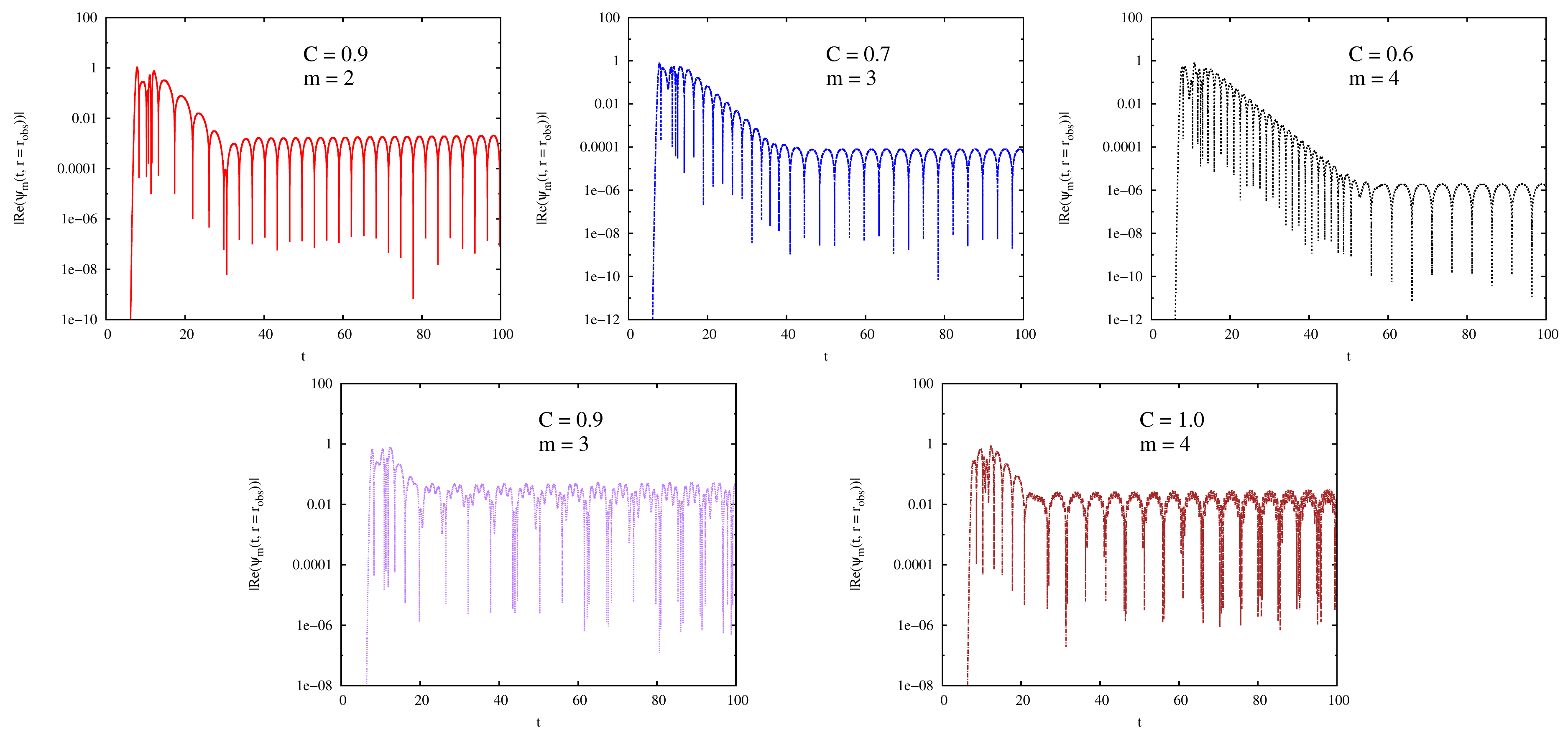}
\caption{Time-domain profiles of $|\text{Re}(\psi_m(t, r))|$ for azimuthal numbers $m=2,\,3,\,4$ and different values of the circulation $C$. Boundary conditions BC I are imposed at $r_{\rm min}=0.3$. We extract the signal at $r_{\rm obs}=10$.}
\label{ring2}
\end{figure*}

Finally, in Fig.~\ref{ring2} we display the interesting behavior at regions of marginal stability. We show in the top plots of Fig.~\ref{ring2}, the time-domain profiles of $|\text{Re}(\psi_m(t, r))|$ for azimuthal numbers $m=2$ (with circulation $C=0.9$), $m=3$ (with circulation $C=0.7$), and $m=4$ (with circulation $C=0.6$). These profiles are dominated by two QNMs with quite different frequencies, and dominating at different stages. 
This is presumably due to the initial data exciting a higher overtone which, because it is strongly damped, eventually gives way to dominance by the fundamental, least damped mode. In the bottom plots of Fig.~\ref{ring2} we show time-domain profiles for azimuthal numbers $m=3$ (with circulation $C=0.9$) and $m=4$ (with circulation $C = 1.0$). These time-domain profiles exhibit a beating pattern presumably due to the presence
of two modes of comparable lifetimes and frequencies. A frequency-domain analysis of these setups does confirm the existence of the two modes showing up in the waveform, which lends further support to the claim that the initial data content is the responsible for the relative excitation of these modes.

\section{Conclusions}
\label{sec-conclusion}

Spacetimes with ergoregions -- regions where negative energy states are accessible -- but without horizons to drain negative energy states, develop
a very generic instability known as ergoregion instability. We have studied in detail this instability for a very appealing effective spacetime describing sound waves in a fluid vortex geometry. Our study makes a direct contact between fluid dynamics and black-hole physics.
Broadbent and Moore have conducted a thorough study of stability of rotating fluids, but imposing different boundary conditions~\cite{Broadbent:1979}. In line with our findings, they uncover an instability for compressible fluids related also to 
sound wave amplification (note that incompressible fluids were also analyzed by Lord Kelvin and were found to be marginally stable~\cite{Kelvin}).

The evidence that the hydrodynamic vortex is an unstable system and that the instabilities are directly related to the existence of an ergoregion
together with the absence of an event horizon agrees with the prediction in Ref.~\cite{Friedman:1978}. This confirmation further strengthens the similarities between effective spacetimes in fluids and black holes.

\begin{acknowledgments}
L. C. and L. O. thank Instituto Superior T\'ecnico (IST), in Lisboa, for kind hospitality during the completion of this work.
The authors would like to thank Conselho Nacional de Desenvolvimento 
Cient\'\i fico e Tecnol\'ogico (CNPq) and Coordena\c{c}\~ao 
de Aperfei\c{c}oamento de Pessoal
de N\'\i vel Superior (CAPES) for partial financial support.
We also acknowledge financial support provided under the European Union's FP7 ERC Starting Grant ``The dynamics of black holes:
testing the limits of Einstein's theory'' grant agreement no. DyBHo--256667 and through the Intra-European Marie Curie contract aStronGR-2011-298297.
This research was supported in part by Perimeter Institute for Theoretical Physics. 
Research at Perimeter Institute is supported by the Government of Canada through 
Industry Canada and by the Province of Ontario through the Ministry of Economic Development 
$\&$ Innovation.
This work was also supported by the NRHEP 295189 FP7-PEOPLE-2011-IRSES Grant, and by FCT-Portugal through projects
CERN/FP/123593/2011.
L. C. is also grateful to the Abdus Salam International Centre for Theoretical Physics through the Associates Scheme.
\end{acknowledgments}



\begin{thebibliography}{99}
\bibitem{drain-footnote} The hydrodynamic vortex can be obtained by considering
the rotating black hole analogue studied in Ref.~\cite{Dolan:2011ti}
(the draining bathtub), by setting the draining to zero ($D=0$),
i.e., when the flow is purely circulating.
\bibitem{huygens_comment} The coordinate $z$ in this work is trivial, and the spacetime under study is effectively three-dimensional.
\end{thebibliography}
\end{document}